\documentclass[twocolumn,showpacs,preprintnumbers,amsmath,amssymb]{revtex4}

\usepackage{graphicx}
\usepackage{dcolumn}
\usepackage{bm}

\begin{document}

\title{Biexcitons in Coupled Quantum Dots as a Source for Entangled Photons}

\author{Oliver Gywat}
\author{Guido Burkard}
\author{Daniel Loss}

\affiliation{Department of Physics and Astronomy, Klingelbergstrasse 82, 
CH-4056 Basel, Switzerland}



\begin{abstract}
We study biexcitonic states in two tunnel-coupled semiconductor
quantum dots and show that such systems provide the possibility to 
produce polarization-entangled photons or spin-entangled electrons 
that are spatially separated at production.
We distinguish between the various spin configurations and calculate
the low-energy biexciton spectrum using the Heitler-London
approximation as a function of magnetic and electric fields.
The oscillator strengths for the biexciton recombination
involving the sequential emission of two photons are calculated.
The entanglement of the polarizations resulting from the spin configuration 
in the biexciton states is quantified as a function of the photon emission 
angles. 
\end{abstract}



\maketitle


Entanglement has been identified as an essential resource for many
applications in the recently developed field of quantum communication
and quantum computation~\cite{qc}.   
Several quantum communication schemes have already been successfully 
implemented with pairs of polarization-entangled photons produced by 
parametric down-conversion~\cite{qc},
 e.g.  the faithful transmission of a quantum state 
(quantum teleportation), entanglement-assisted classical communication 
(e.g., quantum superdense coding), and the production of a secure 
cryptographic key (quantum key distribution).
Recently, there has also been growing interest in solid-state
implementations of quantum computation using the electron spin as
the qubit~\cite{LD}, as well as quantum
communication with spin-entangled electrons.
Superconductor-normal junctions in combination with quantum dots
(QDs) have been suggested as a device for producing entangled 
electrons~\cite{recher}.
Still, the efficient and deterministic production
of both entangled photons and electrons poses 
a theoretical and experimental challenge.
In the case of photons, the use of electron-hole recombination in a single
QD was recently suggested~\cite{benson,moreau}.
Non-resonant excitation of a QD is expected to produce pairs of entangled 
photons with an efficiency (production rate/pump rate) that 
is about four orders of magnitude bigger 
than for parametric down-conversion~\cite{moreau}.

In this paper, we study the production of polarization-entangled photons,
or, alternatively, spin-entangled electrons, using the biexcitonic 
ground state in \textit{two tunnel-coupled} QDs. For this purpose we study
the low-energy biexcitonic states in coupled QDs,
determining their energy spectrum and their optical properties. 
We concentrate on the spin configuration of
the calculated states, being related to the orbital wavefunction via
the Fermi statistics which is implemented in a Heitler-London (HL) 
ansatz for electrons and for holes. 
As a special quality of a double dot,
we find that in the (spin-entangled) biexcitonic ground state,  the 
biexciton favors a configuration with each QD occupied by one exciton, 
thus providing a basis for the separation of the entangled particles.
Even though coupled QDs are usually separated by a distance less than the
wavelength of the
emitted light, it might still be possible to directly detect the photons 
at separate locations. 
It can e.g. be expected that due to 
anisotropies the two dots have different preferred emission 
directions inclosing a fixed angle. Two subsequent photons, which are 
emitted with
a time delay given by the exciton lifetime, could then be detected
separately in the far field.

In contrast to our calculations,
earlier studies for quantum computation or entanglement production with
excitons in QDs concentrate on 
single QDs~\cite{benson,moreau,troiani,chen,steel} 
and/or on charge degrees of freedom 
(neglecting spin)~\cite{troiani,chen,steel,quiroga,bayer,biolatti}.
Also, instead of a pure electrostatic
interdot coupling~\cite{quiroga,biolatti}, we take into account
the tunneling of electrons and holes between the coupled QDs.

Biexcitons consist of two bound excitons which themselves are formed by
a conduction-band electron and a valence-band hole in a semiconductor,
bound together by the attractive Coulomb interaction.
Following the theory of excitonic absorption in single QDs~\cite{efros},
the biexcitonic states in single QDs have been 
investigated~\cite{banyai,takagahara,bryant,hu,nair,hawrylak,kiraz,santori}
and single excitons in coupled QDs
have been observed in experiment~\cite{bayer,abstreiter}.
Recently, spin spectroscopy of excitons in QDs was performed using
polarization-resolved magnetophotoluminenscence~\cite{johnston}.
Two regimes can be distinguished in the
discussion of excitons confined in QDs~\cite{efros}.
In the \textit{weak confinement} 
limit $a_X\ll a_e, a_h$, where $a_X$ is the radius of the free exciton
and $a_e, a_h$ the electron and hole effective Bohr radii in the QD,
an exciton can (as in the bulk material) be considered as a boson
in an external confinement potential. 
In the case of \textit{strong confinement} $a_X\gg a_e, a_h$,
electrons and holes are separately confined in the QD and
the bosonic nature of the electron-hole pair breaks down.
Since, e.g., in bulk GaAs $a_X\approx 10$ nm,
we are in an intermediate regime $a_X\approx a_e, a_h$
for typical QD radii.  
Here, we start from a strong confinement ansatz, i.e.\ from independent
electrons and holes (two of each species), and then use the
HL approximation to include the Coulomb interaction
and the tunneling.
Unlike for bulk biexcitons, where the HL approximation fails for
some values of $\xi=m_{e}/m_{h}$~\cite{brinkman}, we are here in a 
different situation---much more similar to the ${\rm H}_{2}$
molecule---because the single particle orbitals are defined by the
strong QD confinement, the latter playing the role of the
(``infinitely'' heavy) protons of the ${\rm H}_{2}$ molecule.

We obtain the low-energy (spin-resolved) biexciton spectrum
in which the electrons
and holes each form either a spin singlet or triplet.
Subsequently, we calculate the oscillator strength,
being a measure for the optical transition rates.
The spin of the biexciton states relates to
two different states of the polarization-entangled
photon pair produced in the recombination.
We quantify the entanglement of the
photon pair depending on the emission directions.
The variation of the spectrum and the 
oscillator strengths due to 
magnetic or electric fields allows to use
ground-state biexcitons in tunnel-coupled QDs as
a pulsed source of entangled photon pairs.

We model the biexciton (two electrons and two holes)
in two coupled QDs by the Hamiltonian
\begin{equation}
H  =  \sum_{\alpha=e,h}\sum_{i=1}^{2} h_{\alpha i} 
+H_{C}+H_{Z}+H_{E}, \label{h}
\end{equation}
where $h_{\alpha i} = (\mathbf{p}_{\alpha i}+q_{\alpha} \mathbf{A}(\mathbf{r}_{\alpha 
i})/c)^{2}/2m_{\alpha}+V_{\alpha}(\mathbf{r}_{\alpha i})$ is the 
single-particle Hamiltonian for the $i$-th electron
($\alpha=e$, $q_e=-e$) or  hole ($\alpha=h$, $q_h=+e$) 
in two dimensions (2D) with coordinate $\mathbf{r}_{\alpha i}$ and
spin $\mathbf{S}^{\alpha i}$.
The potential 
$V_{\alpha}(x,y)=m_{\alpha}\omega_{\alpha}^{2}[(x^{2}-a^{2})^{2}/4a^{2}+y^{2}]/2$ 
describes two QDs centered at $(x=\pm a, y=0)$, separated by a barrier
of height $m_{\alpha}\omega_{\alpha}^{2} a^2/8$.
Electrons and holes have effective masses $m_{\alpha}$ and
confinement energies $\hbar\omega_{\alpha}$.
The Coulomb interaction is included by 
$H_{C}=(1/2)\sum_{(\alpha,i)\neq (\beta, j)} 
q_{\alpha}q_{\beta}/\kappa |\mathbf{r}_{\alpha i}-\mathbf{r}_{\beta j}|$, 
with a dielectric constant $\kappa$ (for bulk GaAs, $\kappa =13.18$).  
A magnetic field $\mathbf{B}$ in $z$ direction leads to 
orbital effects via the vector potential (in the symmetric gauge)
$\mathbf{A}=B(-y,x,0)/2$ and to the Zeeman term 
$H_{Z}= \sum_{\alpha,i}g_{\alpha}\mu_{B} B S_{z}^{\alpha i}$, 
where $g_{\alpha}$ is the effective g-factor of the electron (hole) 
and $\mu_{B}$ is the Bohr magneton. 
Restricting ourselves to the low-energy physics of QDs filled with 
few particles, we can assume approximately 2D parabolic 
confinement. 
We assume the simultaneous confinement of electrons and holes which
can be realized e.g.\ in QDs
formed by thickness fluctuations in a quantum  well~\cite{steel}
or by self-assembled QDs~\cite{sads,lundstrom}.
A particle in a single QD is thus described by the Fock-Darwin (FD) 
Hamiltonian $h^{\pm a}_{\alpha}(\mathbf{r}_{\alpha i})$~\cite{BLD}, 
comprising a harmonic potential 
$v_{\alpha}^{\pm a} (\mathbf{r}) = m_\alpha\omega_\alpha^2[(x\mp a)^2+y^2]/2$ 
and a perpendicular magnetic field. In prospect of the HL ansatz below we 
write the single-particle part of the Hamiltonian Eq.~(\ref{h}) as 
$\sum_{\alpha}[h^{-a}_{\alpha}({\bf r}_{\alpha 1})
+h^{+a}_{\alpha}(\mathbf{r}_{\alpha 2})]
+H_W(\{\mathbf{r}_{\alpha i}\}) \equiv H_0+H_W$, where $H_W(\{\mathbf{r}_{\alpha i}\})
=\sum_{\alpha}\left[\sum_i V_{\alpha}({\bf r}_{\alpha i}) 
- v^{-a}_{\alpha}({\bf r}_{\alpha 1})- v^{+a}_{\alpha}({\bf r}_{\alpha 2})\right]$. 
An in-plane electric field $\mathbf{E}=\varepsilon {\bf \hat{y}}$
is described by $H_{E}=e\, \varepsilon\,(y_{e1}+y_{e2}-y_{h1}-y_{h2})$ and can be included in $H_0$.
We put $\varepsilon=0$ here and discuss the case $\varepsilon\neq 0$ below.

The valence band is assumed to be split into well-separated heavy and light hole
bands and only heavy-hole excitations are considered in the following.
The FD ground states $|D\rangle_{\alpha}$ in the QD $D=1,2$ which are used to make a
{\em variational} HL ansatz are \cite{BLD}
\begin{equation}
\!\langle {\bf r} |D \rangle_{\alpha}  =  
\sqrt{\!\frac{b_{\alpha}}{\pi a^{2}_{\alpha}}}
\exp{\!\!\left(\!\!-\!\frac{b_{\alpha}}{2a_{\alpha}^{2}}
\left((x\!\pm\!a)^{2}\!+\!y^{2}\right)\!\pm\!\frac{iq_{\alpha}ay}{2el_{B}^{2}}\right)}, 
\label{fds}
\end{equation}
where the upper (lower) sign holds for $D=1\,$($2$),
$l_B=\sqrt{\hbar c/eB}$ and $b_{\alpha}=\sqrt{1+(eB/2cm_{\alpha}\omega_{\alpha})^{2}}$.

\begin{figure}[b]
\centerline{
\includegraphics[width=8.5cm]{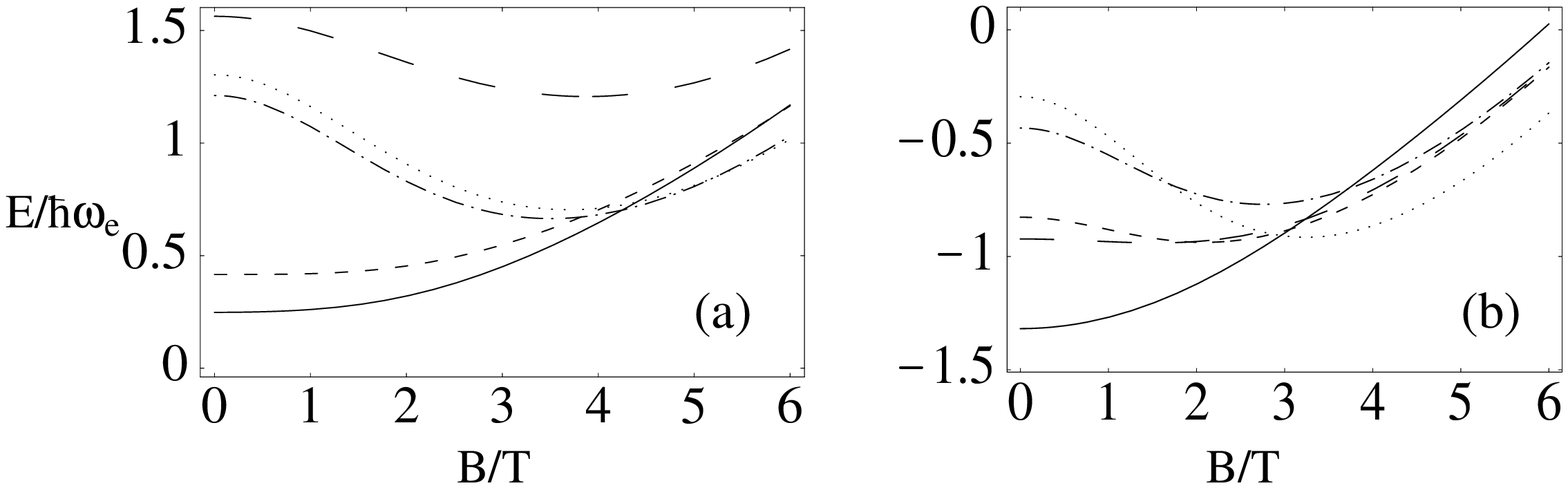}
\vspace{1mm}
}
\caption{Biexciton energies in units of $\hbar\omega_{e}$
for (a) $\eta=\omega_{e}/\omega_{h}=1/2$,
(b) $\eta=1/\xi=1.67$ ($a_e = a_h$),
in a 2D GaAs system
($m_{e}=0.067m_{0}$, $m_{hh}=0.112m_{0}$), $\hbar\omega_{e}=3\;{\rm meV}$,
 and $d=0.7$.
The plotted HL energies $E_{IJ}$ are $E_{ss}$ (solid line), 
$E_{st}$ (short-dashed), $E_{ts}$ (dot-dashed), 
and $E_{tt}$ (dotted), neglecting the
Zeeman energy.
The exchange splittings $E_{tJ}-E_{sJ}$, $J=s,t$,
for electrons are larger than for holes
($E_{It}-E_{Is}$, $I=s,t$)
in (a) where $\eta\xi < 1$, but of the same order in (b) ($\eta\xi = 1$).
At $B=0$, $|ss\rangle$ has the lowest energy, while
for larger $B$, there is a crossover to a $|tt\rangle$ ground state.
Double occupation of a QD (long-dashed line)  becomes more
favorable with increasing $\eta$; 
in (a), $\bar{E}>E_{IJ}$, $I,J=s,t$, while
in (b), $\bar{E}$ is smaller than some of the $E_{IJ}$ for small $B$.}
\label{energies}
\end{figure}

We now make a strong confinement ansatz by constructing
two-particle orbital wave functions for electrons and for holes separately 
according to the HL method, 
i.e. a symmetric ($|s \rangle^{\alpha}\equiv|I=0\rangle^{\alpha}$, 
spin singlet) and an 
antisymmetric ($|t \rangle^{\alpha}\equiv|I=1\rangle^{\alpha}$, spin triplet)
linear combination of two-particle states 
$|DD' \rangle_{\alpha}=|D \rangle_{\alpha}\otimes|D' \rangle_{\alpha}$,
\begin{equation}
|I\rangle^{\alpha} = 
N_{\alpha I}(|12\rangle_{\alpha} +(-1)^{I} |21 \rangle_{\alpha}),
\end{equation}
where $N_{\alpha I}=1/\sqrt{2(1 + (-1)^{I}|S_{\alpha}|^{2})}$ and $S_{\alpha}=\, _{\alpha}\langle 1 | 2 \rangle_{\alpha}$ denotes the overlap (or tunneling amplitude) between the two orbital wave functions $|1 \rangle_{\alpha}$ and $|2 \rangle_{\alpha}$.  We continue by forming the four biexciton states $|IJ \rangle  =  |I\rangle ^{e} \otimes |J\rangle ^{h}$, where $I=0$ ($1$) for the electron singlet (triplet) and $J=0$ ($1$) for the hole singlet (triplet).
The energies
\begin{equation}
E_{IJ}=\langle IJ|H|IJ\rangle = E^{0}+E^{Z}+E_{IJ}^{W}+E_{IJ}^{C},
\end{equation}
with $E_{IJ}^A\equiv \langle IJ|H_A|IJ\rangle$, can be calculated analytically.
In units of $\hbar \omega_{e}$, we find $E_0\equiv E_{IJ}^{0} = 2(b_{e}+b_{h}/\eta)$, 
where $\eta=\omega_{e}/\omega_{h}$, 
$E^{Z}\equiv E_{IJ}^{Z}= (\mu_{B} B/\hbar\omega_{e})\sum_{\alpha i} g_{\alpha}S_z^{\alpha i}$, and
\begin{eqnarray}
 & & E_{IJ}^{W} = \frac{3}{16d^{2}}\left(\frac{1}{b_{e}^{2}}+\frac{\xi}{b_{h}^{2}}\right)-\frac{3d^{2}}{4}\left(1+\frac{1}{\xi\eta^{2}}\right) \nonumber\\
 & &  + 3N_{IJ}\left[d^{2}\left(1+\frac{1}{\xi\eta^{2}}\right)+(-1)^{J}S_{h}^{2}\left(d^{2}-\frac{1}{\eta b_{h}}\right)\right. \nonumber\\
 & &  \left. +(\!-1\!)^{I}S_{e}^{2}\left(\!\frac{d^{2}}{\xi \eta^{2}}\!-\!\frac{1}{b_{e}}\!\right)-(\!-1\!)^{I\!+\!J}S_{e}^{2}S_{h}^{2}\left(\!\frac{1}{b_{e}}\!+\!\frac{1}{\eta b_{h}}\!\right)\right],
\end{eqnarray}
where $2d\!=\!2a/a_{e}$ is the dimensionless inter-dot distance, 
$a_{e}=\sqrt{\hbar/m_{e}\omega_{e}}$ is the electronic Bohr radius,
$S_{e}=\exp{\left(-d^{2}\left[2b_{e}-1/b_{e}\right]\right)}$,
$S_{h}=\exp{\left(-d^{2}\left[2b_{h}-1/b_{h}\right]/\xi\eta\right)}$,
$N_{IJ} = N_{eI}^2N_{hJ}^2$, and $\xi\!\!=\!\!m_{e}/m_{h}$.
For $E_{IJ}^{\rm C}$, we find
\begin{eqnarray}
E_{IJ}^{\rm C} & = & \frac{E_{ee}+(-1)^I\tilde E_{ee}}{1+(-1)^IS_{e}^{2}}+\frac{E_{hh}+(-1)^J\tilde E_{hh}}{1+(-1)^J S_{h}^{2}}\nonumber\\
 & & +8N_{IJ}\left[E_{X}+E_{eh}+(-1)^{I} S_{e} \tilde E_{Xe}\right.\nonumber\\
 & & \left.+(-1)^{J} S_{h} \tilde E_{Xh}+(-1)^{I+J} S_{e}S_{h} \tilde E_{Xeh}\right],
\end{eqnarray}
where we have used the abbreviations 
\begin{eqnarray}
E_{\alpha\alpha} & \!= & \!c\,\sqrt{b_{\alpha}/x_{\alpha}}\exp{\left(-b_{\alpha}d^{2}/x_{\alpha}\right)}\,I_{0}\left(b_{\alpha}d^{2}/x_{\alpha}\right),\\
\tilde E_{\alpha\alpha}  & \!= & \!c\,\sqrt{\frac{b_{\alpha}}{x_{\alpha}}}S_{\alpha}\exp{\!\!\left(\!-\!\frac{b_{\alpha}d^{2}}{x_{\alpha}}\right)}I_{0}\!\left(\frac{d^{2}}{x_{\alpha}}\!\left[b_{\alpha}\!-\!\frac{1}{b_{\alpha}}\right]\right),\\
E_{X} & \!= & \!-c \sqrt{\bar b},\\
E_{eh} & \!= & \!E_{X}\exp{\left(-\bar b d^{2}\right)}\,I_{0}\left(\bar b d^{2}\right),\\
\tilde E_{X\alpha} & \!= & \!2S_{\alpha}E_{X}\exp{\left(-\bar 
bd^{2}/4b_{\alpha}^2\right)}\;I_{0}\left(\bar bd^{2}/4b_{\alpha}^2\right),\\
\tilde E_{Xeh} & \!= & \! S_{e}S_{h}E_{X}\left\{\exp{\left(\bar b_{1}d^{2}/2\right)}I_{0}\left(\bar b_{1}d^{2}/2\right)\right.\nonumber\\
&  & \quad\quad\quad\left.+ \exp{\left(\bar b_{2}d^{2}/2\right)}I_{0}\left(\bar b_{2}d^{2}/2\right)\right\}.
\end{eqnarray}
Here, $I_{0}(x)$ is the zeroth-order modified Bessel function,
$c=e^2\sqrt{\pi/2}\, /\kappa a_{e} \hbar \omega_{e}$ is a dimensionless 
parameter characterizing the Coulomb interaction,
$x_{e}=1$, $x_{h}=\xi\eta$,
$\bar b=2b_{e}b_{h}/(b_{h}+\xi\eta b_{e})$,
$\bar b_{1}=b_{e}-1/b_{e}+\left[b_{h}-1/b_{h}\right]/\xi\eta$,
and $\bar b_{2}=(\left[b_{e}-1/b_{e}\right]\left[b_{h}-2\xi\eta b_{e}\right]+b_{e}\left[b_{h}-1/b_{h}\right])/(b_{h}+\xi\eta b_{e})$. 
Fig.~\ref{energies} shows the biexciton energies $E_{IJ}$ ($I,J=0,1=s,t$)
in the double QD as a function of an applied external magnetic field
in $z$ direction.
The Zeeman interaction $H_Z$ causes an additional level splitting of
$\approx 0.02\, \hbar\omega_e/{\rm T}$ 
(assuming $|g_{e}|\approx |g_{h}|\approx 1$) 
for the triplet states with $\sum_{i}S_z^{\alpha i}\neq 0$
which is not shown in Fig.~\ref{energies}. 
The electron-hole exchange interaction for the GaAs QDs considered here
is reported to be only on the order of tens of
$\mu {\rm eV}$~\cite{gammon} and can therefore be neglected.
The self-consistency of omitting excited single-QD states in the HL 
ansatz can be checked by comparing the energy $E^C_{IJ}+E^W_{IJ}$ 
to the single-QD level 
spacing.
This criterion  is fulfilled for inter-dot distances $2a\gtrsim 20\;\text{nm}$. 
In addition to the HL states $|IJ\rangle$, we consider the double occupation
states $|DDDD\rangle$ for which all four particles are located on the
same QD $D=1,2$.
Their energies are given by $\bar{E} = E^0+E^Z+\bar{E}^{W}+\bar{E}^{C}$,
with $\bar{E}^{W}=3(1/b_{e}^{2}+\xi/b_{h}^{2})/16d^{2}$,
and $\bar{E}^{C}=c(\sqrt{b_{e}}+\sqrt{b_{h}/\xi\eta}-4\sqrt{\bar b})$.

We proceed to the calculation of the oscillator strengths of 
biexciton-exciton and exciton-vacuum transitions. The oscillator 
strength $f$ is a measure for the coupling of exciton states to the 
electromagnetic field and is proportional to the optical transition rates.
For a transition between the $N+1$ and $N$ exciton states $|N+1\rangle$ 
and $|N\rangle$, the oscillator strength is defined as 
\begin{equation}
f_{N+1,N}=2|p_{N {\bf k} \lambda}|^{2}/m_{0}\hbar\omega_{N+1,N}, 
\end{equation}
where $m_{0}$ is the bare electron mass, 
$\hbar\omega_{N+1,N}=E_{N+1}-E_{N}$, and
$p_{N {\bf k} \lambda} = \langle N+1|{\bf e}_{{\bf k} \lambda}\!\cdot {\bf p}|N\rangle$,
where ${\bf e}_{{\bf k} \lambda}$ is the unit polarization vector for a photon 
with momentum ${\bf k}$ and helicity $\lambda =\pm 1$, and ${\bf p}$ is the electron
momentum operator.
For $p_{N \mathbf{k} \lambda}$ we find in the dipole approximation
$a_\alpha\ll 2\pi/k$ ($a_\alpha\approx 20$ nm, $2\pi/k\approx 1$ $\mu$m),
\begin{eqnarray}
 & & p_{N {\bf k} \lambda}  
= [(N\! +\! 1)!]^{2}\!\!\!\!\!
\sum_{\{\sigma_{i},\tau_{j}\},\sigma}\!\!\!\!\!
M_{\sigma\lambda}(\theta)
\int \!\!d^{3}r 
\prod_{i,j}d^{3}r_{i} d^{3}s_{j}
\label{pnkl}\\
& & \times\Phi_{N}(\{{\bf r}_{i},\sigma_{i}\};\{{\bf s}_{j},\tau_{j}\})
\Phi_{N\!+\!1}^{*}(\{{\bf r}_{i},\sigma_{i}\},{\bf r},\sigma;
\{{\bf s}_{j},\tau_{j}\},{\bf r},\sigma\}),
\nonumber
\end{eqnarray}
where $\Phi_{N}$ is the $N$-exciton wavefunction, depending on the conduction-band electron (valence-band hole) coordinates $\mathbf{r}_{i}$ ($\mathbf{s}_{j}$) and their spins $\sigma_{i}$ ($\tau_{j}$) ($i,j=1\dots N$). The coordinate and spin of the electron and the hole created or annihilated during the optical transition are denoted by ${\bf r}$ and $\sigma$.
The inter-band momentum matrix element for a cubic crystal symmetry is given by
$M_{\sigma\lambda}(\theta)={\bf e}_{{\bf k} \lambda} \cdot {\bf p}_{cv}(\sigma) = 
p_{cv}(\cos(\theta)-\sigma\lambda)/2
\equiv p_{cv}m_{\sigma\lambda}(\theta)$,
where $\theta$ is the angle between $\mathbf{k}$ and the
normal to the plane of the 2D electron system
(assuming that the latter coincides with
one of the main axes of the cubic crystal), 
and $E_p=2p_{cv}^2/m_0$ ($=25.7\,{\rm eV}$ for GaAs).

\begin{figure}[b]
\centerline{
\includegraphics[width=8.5cm]{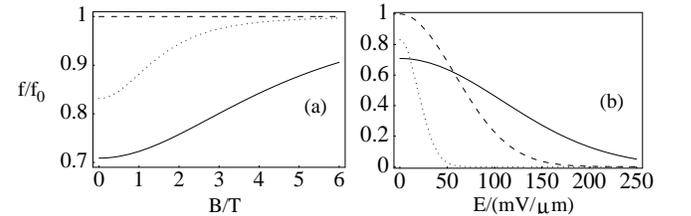}
\vspace{1mm}
} 
\caption{Oscillator strength $f_{X,0}$ for GaAs QDs in units of $f_{0}$ 
as a function of (a) the magnetic field $B$ (in Tesla) at $E=0$ and 
(b) the electric field $E$ (in ${\rm mV/}\mu {\rm m}$) at $B=0$, 
with $\eta=\omega_{e}/\omega_{h}=1/2$ (solid line), $\eta=1/\xi$ 
(dashed), $\eta=4$ (dotted). For $\eta=1/\xi$ the $B$ field has no effect
on $C_{eh}$.}
\label{ceh}
\end{figure}

According to Eq.~(\ref{pnkl}), the orbital momentum matrix element for 
transitions from the exciton vacuum $|0\rangle$ to an exciton state 
$|X\rangle = |D\rangle _e\otimes |D\rangle _h\equiv |DD\rangle$ in one QD (or for the
optical recombination of $|X\rangle$) is 
$ p_{0} =  M_{\sigma\lambda}(\theta)\int \!\!d^{3}r \Phi^{*}_{1}({\bf r},{\bf r})\equiv M_{\sigma\lambda}(\theta) C_{eh}$.
The exciton wave function is
denoted by $\Phi_{1}({\bf r}_{e},{\bf r}_{h})=\langle {\bf r}_e,{\bf r}_h|X\rangle$.
From this, we find for the oscillator strength
\begin{equation}
f_{X,0}=\frac{2 |p_{0}|^2}{m_{0}\hbar \omega_{X,0}}
=\frac{E_{p}}{\hbar \omega_{X,0}}M_{\sigma\lambda}(\theta)^2|C_{eh}|^{2},
\end{equation}
and $C_{eh} =2\sqrt{\xi\eta b_{e}b_{h}}/(b_{h}+\xi\eta b_{e})$.
In Fig.~\ref{ceh}a we plot $|C_{eh}|^{2}=f/f_0$ as a function of the magnetic field,
where $f_{0}=E_p m_{\sigma\lambda}(\theta)^2/E_g$ denotes the oscillator strength for (bulk) 
inter-band transitions, equating $\hbar\omega_{X,0}$ with the band-gap 
energy $E_g$. Since we have made a strong confinement ansatz, the obtained oscillator strength is independent of the QD volume $V$. For weak confinement, one would expect $f\propto V$. Fig.~\ref{ceh}b shows 
the suppression of the exciton transition rate by an electric field.

The momentum matrix element $p_1$ for transitions from an exciton
state $|X\rangle$ to a biexciton state $|XX\rangle$ is given by
$p_{1} =  -2 M_{\sigma\lambda}(\theta)\int \!\!d^{3}r_{e} d^{3}r_{h} d^{3}r 
\Phi^{*}_{2}({\bf r}_{e},{\bf r};{\bf r}_{h},{\bf r})
\Phi_{1}({\bf r}_{e},{\bf r}_{h})$.
If the recombining electron and hole are on the same QD, the integral over ${\bf r}$ yields $C_{eh}$, otherwise $S_{eh} = C_{eh} \exp{\left(-2d^{2}\left[b_{e}-\xi\eta/(b_{h}+\xi\eta b_{e})\right]\right)}$.

\begin{figure}[h]
\centerline{
\includegraphics[width=8.5cm]{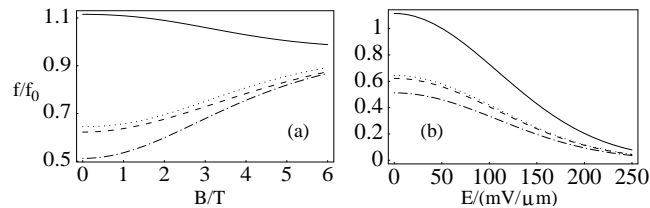}
\vspace{1mm}
}
\caption{Oscillator strengths $f_{XX,X}$
for transitions between the biexciton states $|XX\rangle =|IJ\rangle$ 
and a single remaining
exciton on one QD in units of $f_0$  as a function of
(a) the magnetic field $B$ (in Tesla) at $E=0$ and 
(b) the electric field $E$ (in ${\rm mV/}\mu {\rm m}$) at $B=0$.
The parameters were chosen for GaAs with $\eta=\omega_{e}/\omega_{h}=1/2$.
The line styles correspond to those for $E_{IJ}$ in Fig.~\ref{energies}.}
\label{os1}
\end{figure}

We give here our result for $p_{1}$ for a
transition between the HL biexciton states $|XX\rangle = |IJ\rangle$ with one 
exciton on each QD and a single exciton in the final state 
$|X\rangle = |DD\rangle$, a single exciton on dot $D=1,2$,
\begin{eqnarray}
 & & |\langle IJ|{\bf e}_{{\bf k}\lambda}\!\cdot{\bf p}|DD\rangle | 
  =  2 M_{\sigma\lambda}(\theta) \sqrt{N_{IJ}}\nonumber\\
& &\times\left(C_{eh}\left[(\!-1\!)^{I\!+\!J}\!\!\!+\!S_{e}S_{h}\right]
+S_{eh}\left[(\!-1\!)^{J}\!S_{e}\!+\!(\!-1\!)^{I}\!S_{h}\right]\right).
\label{DD}
\end{eqnarray}
Approximating  $\hbar\omega_{XX,X}\approx E_{g}$, we plot the 
corresponding oscillator strength versus $B$ and $E$ in Figs.~\ref{os1}a 
and~\ref{os1}b.

Results for $f_{XX,X}$, also including the (biexciton) double occupation state $|DDDD\rangle$ and various final (exciton) states, will be given 
elsewhere~\cite{unpublished}.

The main effect of an electric field 
is to spatially separate the electrons from the holes~\cite{tune},
which leads to a reduction of the oscillator strengths~\cite{unpublished} 
(cf. Figs.~\ref{ceh}b and~\ref{os1}b).
Hence, the optical transition rate can be efficiently switched off and on, thus allowing the deterministic emission of one photon pair.

Transformation of a HL biexciton state $|IJ\rangle$ into the basis of two 
coupled excitons yields a superposition of dark 
($S_{z}=\pm 2$) and bright ($S_{z}=\pm 1$) exciton states.
The emitted photon states are (up to normalization)
\begin{equation}
|\chi_{IJ}\rangle \propto
|\! +\!1,\theta_{1}\rangle|\! -\! 1,\theta_{2}\rangle + 
(-1)^{I\! +\! J}|\! -\! 1,\theta_{1}\rangle|\! +\!1,\theta_{2}\rangle,
\label{photons}
\end{equation}
where $|\sigma,\theta\rangle
=N(\theta)(m_{\sigma,+1}(\theta)|\sigma_{+}\rangle 
+m_{\sigma,-1}(\theta)|\sigma_{-}\rangle)$
is the state of a photon emitted from the recombination of
an electron  with spin $S_{z}=\sigma/2=\pm 1/2$ and a heavy hole with
spin $S_{z}=3\sigma/2$ in a direction which
encloses the angle $\theta$ with the normal to the plane of the
2D electron and hole motion.
The states of right and left circular
polarization are denoted $|\sigma_{\pm}\rangle$. 

The entanglement of the state~(\ref{photons}) can be
quantified by the von Neumann entropy $E$. For $|ss\rangle$ or $|tt\rangle$ 
and emission of the two photons enclosing an azimuthal angle $\phi=0$ or $\pi$, we obtain $E=\log_{2}(1+x_{1}x_{2})-x_{1}x_{2}\log_{2}(x_{1}x_{2})/(1+x_{1}x_{2})$,
where $x_{i}=\cos^2(\theta_{i})$.
Note that only the
emission of both photons perpendicular to the plane 
($\theta_{1}=\theta_{2}=0$)
results in maximal entanglement ($E=1$) since only in this case
$|\! +\!1,\theta_i\rangle$ is orthogonal to $|\! -\! 1,\theta_i\rangle$.
In particular, the two photons are not entangled ($E=0$) if at least
one of them is emitted in-plane ($\theta_{i}=\pi/2$).
To observe the proposed effect, the relaxation rate to the biexciton 
ground state must exceed the biexciton recombination rate.
That such a regime can be reached is suggested by experiments with low 
excitation densities, see e.g.~\cite{ohnesorge}.
Then, an upper limit for the pair production rate is given by 
$(\tau_{X}+\tau_{XX})^{-1}$, where $\tau_{X,XX}$ is the (bi)exciton lifetime.

Conversely, spin-entangled electrons can be produced 
by optical absorption followed by relaxation of the biexciton to its 
ground state.
After each QD has been filled with an exciton, the
recombination can be suppressed by an
electric field.
Having removed the holes, the electron singlet and triplet
could then in principle be distinguished by a subsequent
interference experiment~\cite{noise}.

We thank A.~V. Khaetskii, A. Imamo\=glu, and P. Petroff for discussions.
We acknowledge support from the Swiss NSF, DARPA, and ARO.



\begin{thebibliography}{9}
\bibitem{qc}
C.H.\ Bennett, D.P.\ DiVincenzo, Nature {\bf 404}, 247 (2000).

\bibitem{LD}
D. Loss, D.~P.\ DiVincenzo, Phys.\ Rev.\ A {\bf 57}, 120 (1998).

\bibitem{recher}
P. Recher, E.~V. Sukhorukov, D. Loss,
Phys.\ Rev.\ B {\bf 63}, 165314 (2001).

\bibitem{benson}
O. Benson \textit{et al.}, Phys. Rev. Lett. ${\bf 84}$, 2513 (2000).

\bibitem{moreau}
E. Moreau \textit{et al.}, Phys. Rev. Lett. ${\bf 87}$, 183601 (2001).

\bibitem{troiani}
F. Troiani, U. Hohenester, E. Molinari,
Phys.\ Rev.\ B ${\bf 62}$, R2263 (2000);

\bibitem{chen}
P. Chen, C. Piermarocchi, L.~J. Sham,
Phys.\ Rev.\ Lett.\ {\bf 87}, 067401 (2001).

\bibitem{steel}
G. Chen \textit{et al.}, Science {\bf 289}, 1906 (2000).

\bibitem{quiroga}
L. Quiroga, N.~F. Johnson, Phys.\ Rev.\ Lett.\ {\bf 83}, 2270 (1999).

\bibitem{bayer}
M. Bayer \textit{et al.}, Science {\bf 291}, 451 (2001).

\bibitem{biolatti}
E. Biolatti \textit{et al.},  Phys.\ Rev.\ B ${\bf 65}$, 075306 (2002).

\bibitem{efros}
Al.~L. Efros, A.~L. Efros,
Sov.\ Phys.\ Semicond.\ {\bf 16}, 772 (1982).

\bibitem{banyai}
L. Banyai \textit{et al.}, Phys.\ Rev.\ B ${\bf 38}$, 8142 (1988).

\bibitem{takagahara} 
T. Takagahara,  Phys. Rev. B ${\bf39}$, 10206 (1989).

\bibitem{bryant}
G.~W. Bryant, Phys.\  Rev.\ B ${\bf 41}$, 1243 (1990).

\bibitem{hu} 
Y.~Z. Hu \textit{et al.}, Phys.\  Rev.\ Lett.\ ${\bf 64}$, 1805 (1990);\\
Phys.\  Rev.\ B ${\bf 42}$, 1713 (1990).

\bibitem{nair}
S.~V. Nair, T. Takagahara, Phys. Rev. B {\bf 55}, 5153 (1996).

\bibitem{hawrylak}
P. Hawrylak,  Phys. Rev. B {\bf 60}, 5597 (1999).

\bibitem{kiraz}
A. Kiraz \textit{et al.}, cond-mat/0108450.

\bibitem{santori}
C. Santori \textit{et al.}, cond-mat/0108466.

\bibitem{abstreiter}
G. Schedelbeck \textit{et al.}, Science {\bf 278}, 1792 (1997).

\bibitem{johnston}
E. Johnston-Halperin \textit{et al.},
Phys.\ Rev.\ B {\bf 63}, 205309 (2001).

\bibitem{brinkman}
W.~F. Brinkman, T.~M. Rice, B. Bell, Phys. Rev. B {\bf 8}, 1570 (1973).

\bibitem{sads}
R.~J. Luyken \textit{et al.},
Physica E {\bf 2}, 704 (1998).

\bibitem{lundstrom}
T. Lundstrom \textit{et al.}, Science {\bf 286}, 2312 (1999).


\bibitem{BLD}
G. Burkard, D. Loss, D.~P. DiVincenzo,
Phys.\ Rev.\ B ${\bf59}$, 2070 (1999).

\bibitem{gammon}
D. Gammon \textit{et al.}, Science {\bf 273}, 87 (1996).

\bibitem{unpublished}
O. Gywat, G. Burkard, D. Loss, to be published.


\bibitem{tune}
For non-identical dots, electric fields can also be used to 
tune the lowest electron levels into resonance which
is sufficient for the generation of entangled photons or electrons.

\bibitem{ohnesorge}
B. Ohnesorge \textit{et al.}, Phys. Rev. B {\bf 54}, 11532 (1996).

\bibitem{noise}
G. Burkard, D. Loss, E.~V. Sukhorukov,  Phys. Rev. B {\bf 61}, R16303 (2000).

\end{thebibliography}
\end{document}